\documentclass{article}

\usepackage{arxiv}

\usepackage[utf8]{inputenc} % allow utf-8 input
\usepackage[T1]{fontenc}    % use 8-bit T1 fonts
\usepackage{hyperref}       % hyperlinks
\hypersetup{
    colorlinks,
    linkcolor={red!50!black},
    citecolor={blue!50!black},
    urlcolor={blue!80!black}
}
\usepackage{url}            % simple URL typesetting
\usepackage{booktabs}       % professional-quality tables
\usepackage{amsfonts}       % blackboard math symbols
\usepackage{nicefrac}       % compact symbols for 1/2, etc.
\usepackage{microtype}      % microtypography
\usepackage{lipsum}		% Can be removed after putting your text content
\usepackage{graphicx}
\usepackage{doi}

\usepackage{lineno}
\usepackage{wrapfig}
\usepackage{siunitx}
\usepackage{svg}
\usepackage{subcaption}

\DeclareSIUnit{\pixel}{pixel}

\title{ALICE ITS3: a bent stitched \mbox{MAPS-based vertex detector}}

\date{November 3, 2023}	% Here you can change the date presented in the paper title
\author{ \href{https://orcid.org/0000-0003-0761-7401}{\includegraphics[scale=0.06]{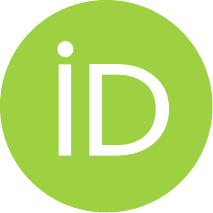}\hspace{1mm}Ola Groettvik} \\
	European Organization for Nuclear Research (CERN)\\
    Geneva, Switzerland \\
	\texttt{ola.groettvik@cern.ch} \\\\
    \textbf{On behalf of the ALICE Collaboration}
}

\renewcommand{\shorttitle}{}

\hypersetup{
pdftitle={ALICE ITS3: A bent stiched MAPS-based vertex detector},
pdfauthor={Ola Groettvik},
}

\begin{document}
\maketitle

\begin{abstract}
	The ALICE ITS3 is a novel vertex detector replacing the innermost layers of ITS2 during LS3.
Composed of three truly cylindrical layers of wafer-sized 65 nm stitched Monolithic Active Pixel Sensors,
ITS3 provides high-resolution tracking of charged particles generated in heavy-ion collisions.
This contribution presents an overview of the ITS3 detector, highlighting its design features,
integration and cooling, and the ongoing development towards the final sensor.
Furthermore, the paper introduces the off-detector service electronics, which play an essential
role in the readout, control, and power supply of the detector.
\end{abstract}

% keywords can be removed
\section{Introduction}

The Inner Tracking System (ITS) is a key component of ALICE, located at the very heart of the detector setup, surrounding the interaction point and providing precise tracking of charged particles produced in heavy-ion collisions. The second iteration,  ITS2, was installed before the start of LHC RUN 3 and consists of 7 layers of roughly 24k CMOS Monolithic Active Pixel Sensors (MAPS). We differentiate between the Inner Barrel (IB), the 3 innermost layers, and the Outer Barrel (OB), the 4 outer layers. ITS3, the third version of ITS, will replace the IB of ITS2 during the LHC Long Shutdown 3 (2026-28).

The ITS3 aims to improve the ITS2 detector on two key parameters: \textcolor{black}{1) by reducing the material budget and 2) by reducing the distance to the interaction point.}

\section{Improving the ITS2 Detector}\label{sec:its2}

\textcolor{black}{ITS2 is a densely packed detector.} Each ITS2 IB stave is made up of 9 silicon sensors. However, the stave also includes a \textcolor{black}{flexible printed circuit board (FPC)} with aluminum for power distribution and signal routing and kapton as a dielectric, a cold plate for water cooling and carbon fiber mechanical structures that hold everything together. The stave average material budget is $\qty{0.35}{\percent}~X_0$ \cite{a}.
The sensitve material of the detector, the silicon, makes up only 15 \% of this material budget. In addition, as a result of irregularities due to the cooling and mechanical support structures, the material peaks at $\qty{0.5}{\percent}~X_0$.

The extra material degrades the pointing resolution, while, the higher the pointing resolution is, the better one can resolve secondary vertices from the decay of short-lived particles containing charm and beauty quarks. To significantly improve the pointing resolution, the ITS3 aims to remove all the extra material that is not actively used for detecting and tracking particles, i.e., resulting in a detector that consists almost exclusively of silicon.

To do this, one must, first of all, replace the water cooling with air cooling. Secondly, by having the data and control signals and power distribution integrated on the sensor silicon, one can remove the external circuit board. Finally, one can remove most of the mechanical support by making the sensor have a self-supporting arched structure.

\textcolor{black}{Furthermore, the innermost layer of ITS2 is \qty{24}{\milli\metre} from the interaction point. For ITS3, the radial position of the first layer is reduced by \qty{6}{\milli\metre}, i.e. to \qty{18}{\milli\metre} (see Figure~\ref{fig:its3}).}

\textcolor{black}{Because of this reduction of material and radial distance to the interaction point, the ITS3 detector improve the spatial resolution with a factor of 2 compared to the one of the current ITS2 detector.}

\section{The ITS3 Detector}

\begin{figure}[h]
  \vspace*{-8mm}
  \begin{center}
    \includegraphics[width=0.54\textwidth]{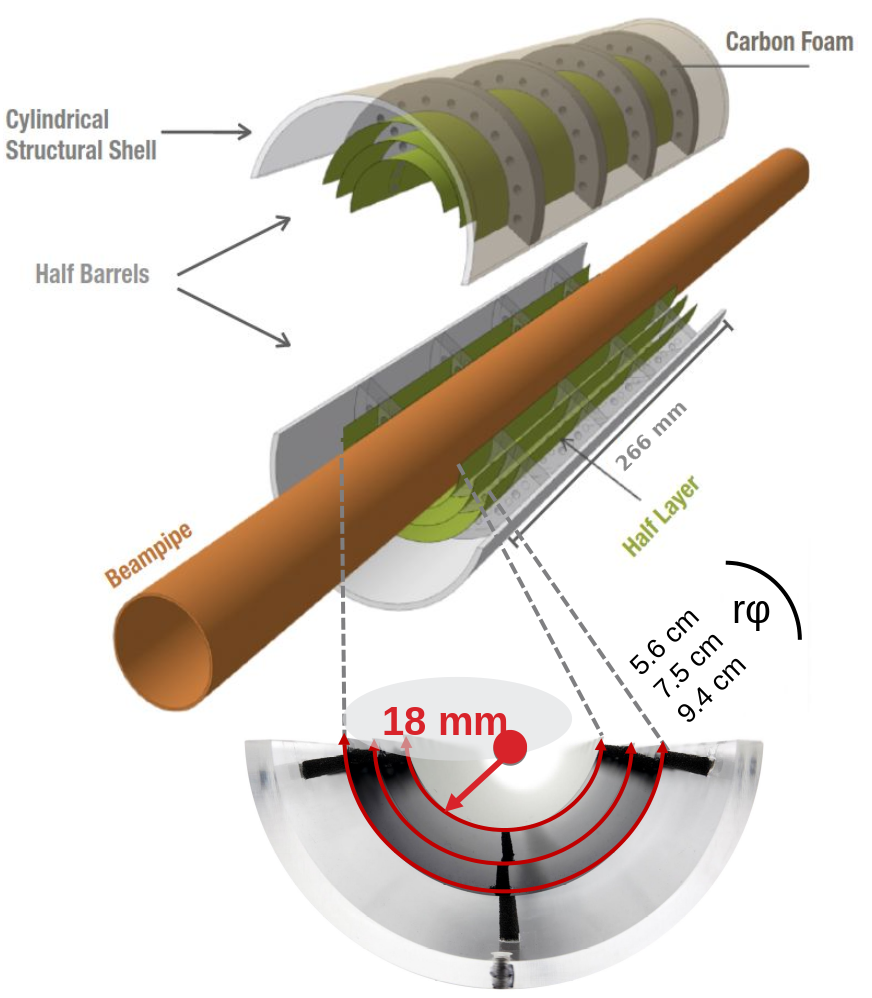}
  \end{center}
  %\vspace*{-7mm}
  \caption{\label{fig:its3} The ITS3 detector concept shown as a drawn schematic layout (top) and one of several mechanical models developed (bottom). }
\end{figure}

To achieve all this, the ITS3 detector uses wafer-scale sensor ASICs, where each half-layer is constructed as a complete piece of silicon, as illustrated with the green sheets in Figure~\ref{fig:its3}. Stitching is used to manufacture a sensor that is larger than the limitations of the lithographic equipment. It allows electrical connections between design units which size is limited by the maximum reticle size. Thus, routing of power distribution and signals can be achieved on the sensor chips themselves, avoiding any external circuit boards.

The sensors are ultra-thin at \qty{50}{\micro\metre} and thus can be bent around the beam pipe. Notice that because of the different radial distances to the interaction point, the various layers are bent at different radii, and are therefore slightly different in width, i.e. \unit{r\varphi}, to make up a full half-layer. The widths are \qty{56}{\milli\metre}, \qty{75}{\milli\metre} and \qty{94}{\milli\metre} for layer 0 to layer 2, respectively. The length in the beam direction, z, is \qty{266}{\milli\metre}, and is the same for all layers.

The ITS3 sensor is a MAPS using 65 nm CMOS technology, a change from the 180 nm CMOS technology of the ITS2 sensor. This change does not only allow for denser circuitry, but is also crucial to create large enough sensors since this particular process is manufactured on larger wafers. Notice that in Figure~\ref{fig:wafer}, the length of the sensors in the z-direction is close to the full diameter of the wafer.

The ITS3 detector will be extremely lightweight, with a material budget reduction of \qty{80}{\percent} to $\qty{0.05}{\percent}~X_0$. The material will be uniformly distributed, except for very light carbon fiber support structures and glue in strategic locations.

\section{Air-cooling}

A custom wind-tunnel setup is built and used together with breadboard models of the detector to study the efficiency of air-cooling and its impact on the detector vibration. Detector vibration has a potential negative impact on the spatial resolution, given that the position of the pixel structures will briefly shift. The airflow's contribution to the vibration is measured to be within  \qty{\pm0.5}{\micro\metre}, and will have a negligible effect on the resolution.

Open-cell carbon foam spacers act as radiators and are used to increase the surface contact with the airflow and provide structural support. With the current power estimation, i.e. roughly \qty{40}{\milli\watt\per\centi\metre\squared}, providing an airflow of 20 degrees at \qty{8}{\metre\per\second} is sufficient to keep the sensor at the nominal operating temperature \textcolor{black}{of} \qty{25}{\celsius}. This is verified with both simulations and by measurements using the wind-tunnel test setup.

\section{Towards the Final Sensor: MOSAIX}

\begin{figure}[h]
  \centering % \begin{center}/\end{center} takes some additional vertical space
  \includegraphics[width=.97\textwidth, trim=0 0 0 77, clip]{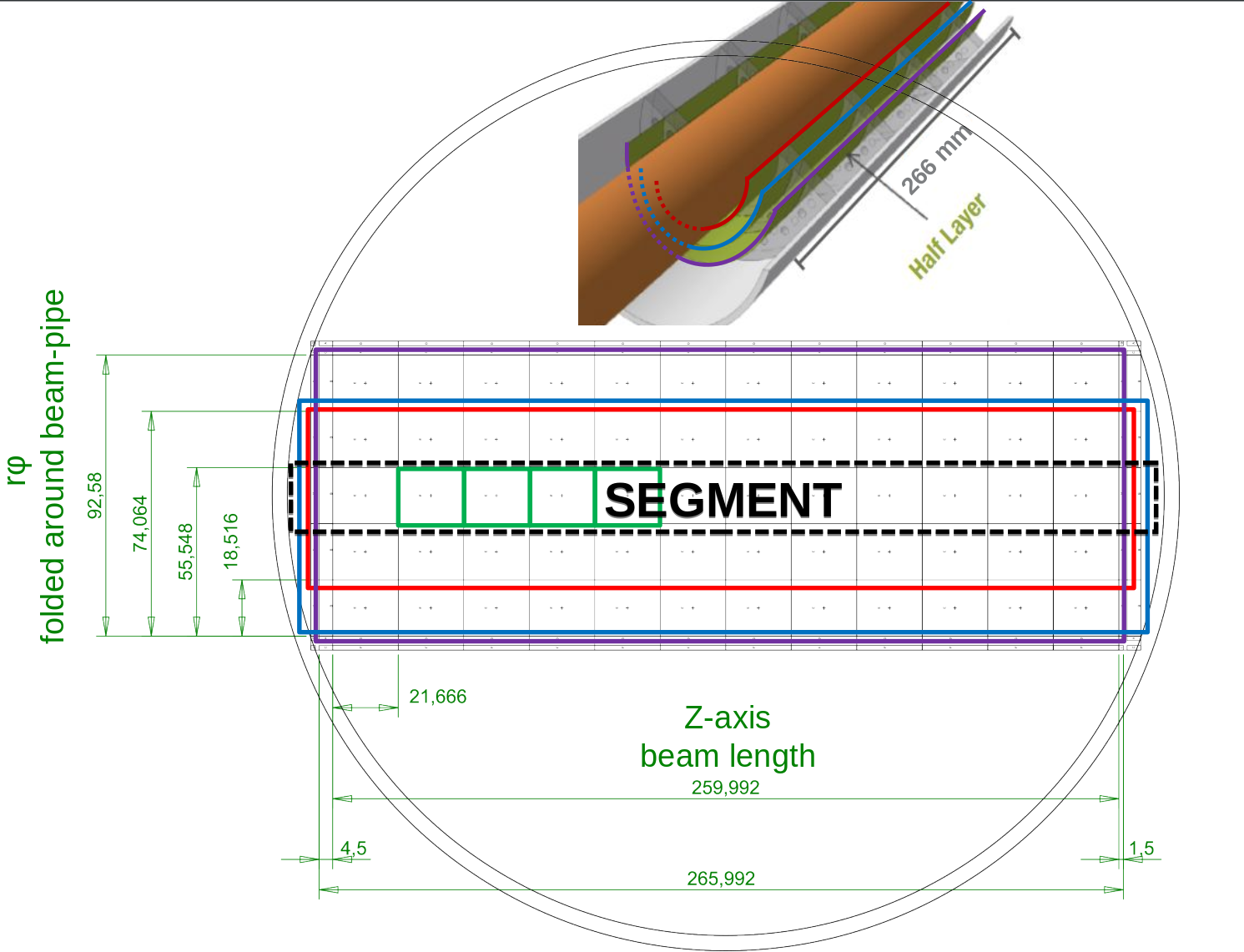}
  \caption{\label{fig:wafer} Wafer floorplan of the MOSAIX sensor. The outline of half-layer 0, 1 and 2 are shown in red, blue and violet, respectively. The green rectangles shows the location of a subset of the Repeatable Sensor Units. The outline of a segment is shown with a black stipled rectangle.}
\end{figure}

The sensor R\&D efforts started by studying the effects of bending MAPS and it has been shown that detection efficiency is agnostic to the bending radius~\cite{b}. Secondly, a 65 nm CMOS technology process has been verified for the ITS3 application by the developing and qualifying analog and digital test structures. The design parameters and operational range that keeps the detection efficiency above \qty{99}{\percent} and the fake-hit rate below \qty{0.01}{\per\pixel\per\second} have been identified~\cite{c}. In addition, the technology has been verified in terms of radiation hardness. A first full sensor prototype that applies stitching, the MOSS, has been produced and is currently being tested to assess yield and performance.

The sensor prototype, the MOSAIX, is currently being designed. Figure~\ref{fig:wafer} shows the wafer floorplan. The outlines of each half-layer sensor are shown in different colors and are mapped to the actual location in the detector in the figure above, showing how they are bent around the beampipe. The sensor is mainly composed of Repeatable Sensor Units (RSU), a subset of them are shown as green rectangles. The size of the RSUs is at the upper limit of the lithographic reticle. Thus, all electrical connections between them are accomplished via a stitched backbone in the z-direction. There are no electrical connections between the RSUs in the \unit{r\varphi} direction.

Thus, naturally independent units called segments appear, i.e. each row of RSUs on the half-layers. These can be considered to replace the ITS2 independent staves, even though they are indeed part of the same piece of silicon. Notice that half-layer 1 and 2, are respectively 1 and 2 segments wider than half-layer 0. Each segment consists of 3 distinct parts or sub-designs. Power is supplied to both the Right End-Cap (REC) and the Left End-Cap (LEC). However, on the LEC, the I/O is connected for the control and data transmission of the sensor. In addition, logic to encode the data for high-speed data transmission is located on the LEC. Finally, each segment consists of 12 instances of the RSU, which is the main unit with the pixel matrix.

The data are transmitted from all the units, all the way to the left side of the segment. In addition, note that the power is distributed on the stitched backbone from both the segment edges and to the middle. \textcolor{black}{This is done to counteract the ohmic potential drop, i.e. the IR drop, which reduces the supply voltage and must be kept at a maximum \qty{10}{\percent} of the input voltage. The drop goes quadratically with the length of the sensor and the amount of available metal is shared 50-50 between supply and return paths. Supplying from both sides thus reduces the drop by a factor of 4. The target process does not provide a metal layer with a sheet resistance low enough to allow for a supply from a single side (<\qty{5.3}{\milli\ohm}), while the requirement is in reach for a two-side supply given a thick metal layer providing a sheet resistance <\qty{10}{\milli\ohm}.}

Each RSU is split in a top and bottom mirrored parts, and then further divided into 6 tiles. Each tile can be thought of as independent sensors, each with its own controllable power domain. In case of shorts or other issues, a tile can be turned off without affecting the operation of neighbouring tiles. The data from the pixel matrix are transferred from the tile on independent links at 160 Mbps. Rebuffering stages are added along the z-axis to ensure that the data are transmitted all the way to the LEC, where the data from the 144 tile links are combined before it is offloaded on high-speed links.

\section{Services and Integration}

\begin{figure}[h]
  \vspace*{-8mm}
  \begin{center}
    \includegraphics[width=0.54\textwidth]{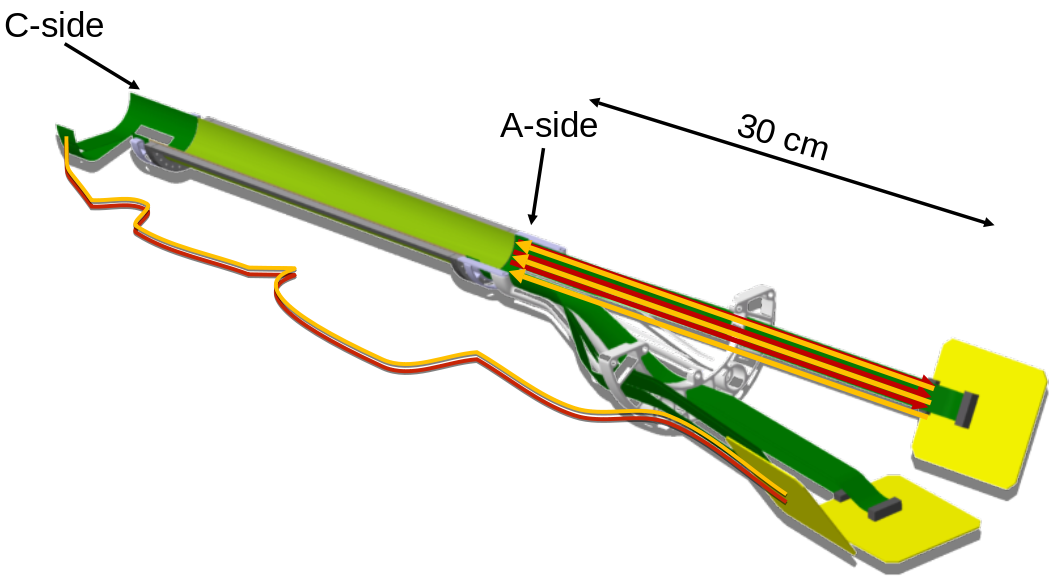}
  \end{center}
  \caption{\label{fig:integration} The sensor integration with the service electronics.}
\end{figure}

Each sensor half-layer is wire-bonded on both sides to FPCs. The main FPC on the A-side provides the I/O traces to the sensor, i.e. it is connected to the sensor LEC. As seen in Figure~\ref{fig:integration}, the FPCs are connected to service electronics PCBs, that make use of recent CERN developments for both power control and data transmission.
% Here, the electrical signals are converted to optical signals, and there are DC-DC regulators for each power domain of the sensor segment.

In the ITS2 services architecture the readout boards are connected to the detector via \qty{8}{\metre} long cables. This is excluded with the the ITS3 detector, as data is transmitted off the sensor at \qty{10.24}{\giga\bit\per\second}. Driving the electrical signals at that rate is not feasible at such distances without inserting active repeaters in the data path. Instead, for the ITS3 services, we opt to convert the electrical signals to optical signals as early as possible, within a maximum of \qty{50}{\centi\metre}.

\begin{figure}[h]
  \centering % \begin{center}/\end{center} takes some additional vertical space
  \begin{subfigure}{0.5\textwidth}
    \centering % \begin{center}/\end{center} takes some additional vertical space
    \includegraphics[width=1.4\textwidth, angle=90]{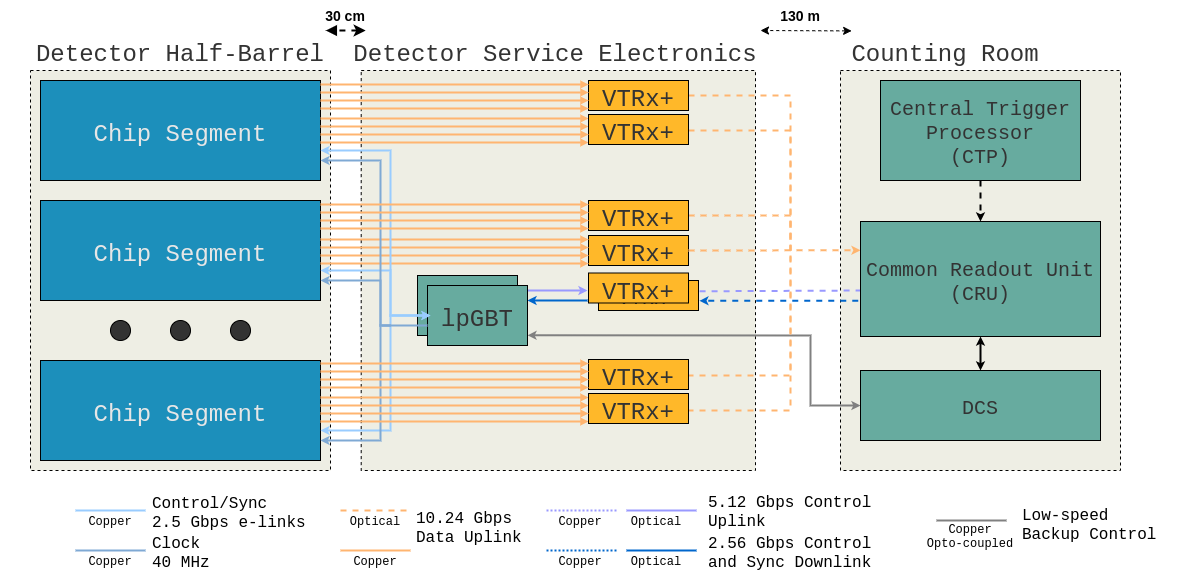}
    \caption{Data and control.}\label{fig:services_data}
  \end{subfigure}%
  \begin{subfigure}{0.5\textwidth}
    \centering % \begin{center}/\end{center} takes some additional vertical space
    \includegraphics[width=1.4\textwidth, angle=90]{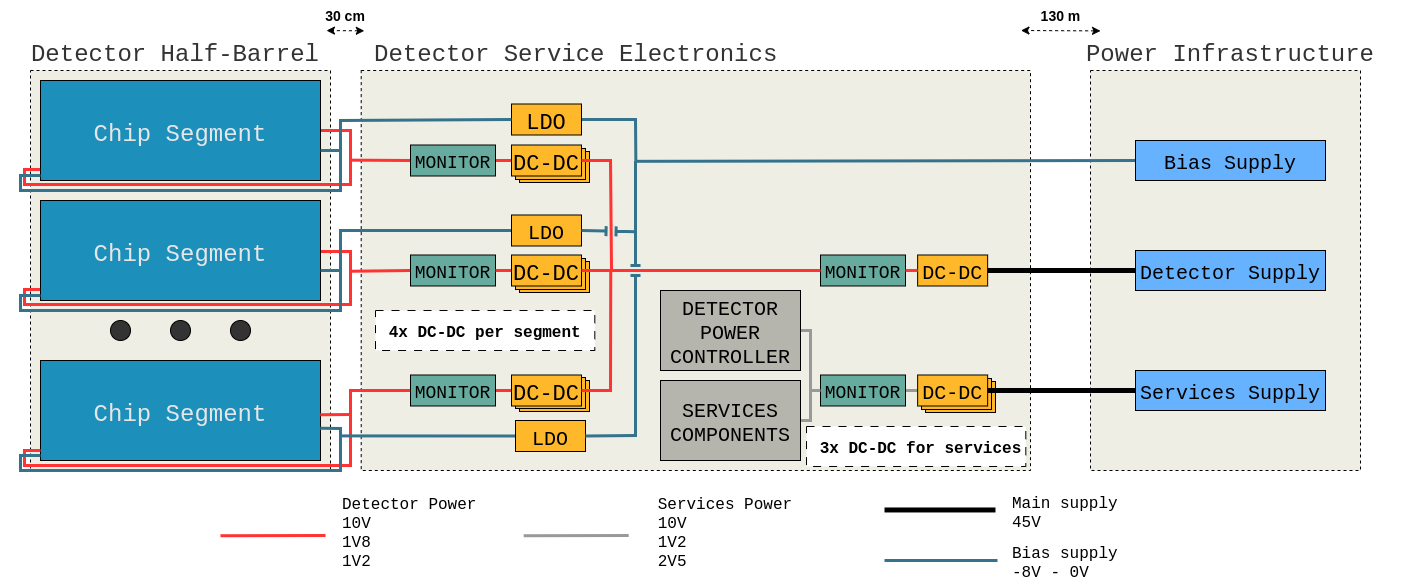}
    \caption{Power.}\label{fig:services_power}
  \end{subfigure}%
  \caption{\label{fig:services} The ITS3 service electronics architecture.}
\end{figure}

To ship the data from the sensor to the upstream ALICE infrastructure the services make use of recent developments from the \textcolor{black}{community; such as an} IP core from the CERN-developed lpGBT for encoding the data on the sensor~\cite{d}. As seen in Figure~\ref{fig:services_data}, this makes it possible to directly drive the radiation hard optical transceiver, the VTRx+~\cite{e}, from the sensor ASIC. The VTRx+ is also a recent CERN-developed component. Fiber optical links are connected to the ALICE experiment Common Readout Units, which process the data and transmit it further for online and offline data processing. The direct translation to an optical signal greatly reduces the complexity of the readout system compared to the ITS2 readout system, which has complex FPGA-based readout boards in the readout path in the radiation environment. Crucially, in this scheme, no FPGAs are placed in the radiation environment.

The Common Readout Unit is also used to transmit and receive the slow control communication messages and transmit trigger commands. In the control scheme, the lpGBT ASICs themselves are used to distribute the signals to the sensor segments via e-links.

The service electronics are also responsible for powering the detector. The CERN-developed radiation tolerant DC-DC converters, bPOLs, are employed for this purpose, as seen in Figure~\ref{fig:services_power}.

\section{Summary}

The ITS3 detector will replace the Inner Barrel of ITS2 during the third LHC long shutdown. This paper summarized the ITS3 detector key design features and how it will be integrated in the ALICE experiment. The detector makes use of wafer-scale sensors to drastically reduce the material budget and to reduce the radial distance from the interaction point by 1) replacing water cooling with air cooling, 2) route electrical signal on-chip and 3) by having a self-supporting arched structure. An overview of the next sensor, the MOSAIX, has been presented. Finally, we have presented how late CERN-developments like the lpGBT, VTRx+ and bPOLs have greatly simplified the architecture of the integration and services as compared to the ITS2 detector.


\end{document}